\documentclass[sigconf]{acmart}
\copyrightyear{2024}
\acmYear{2024}
\setcopyright{acmlicensed}
\acmConference[ESEM '24]{Proceedings of the 18th ACM / IEEE International Symposium on Empirical Software Engineering and Measurement}{October 24--25, 2024}{Barcelona, Spain}
\acmBooktitle{Proceedings of the 18th ACM / IEEE International Symposium on Empirical Software Engineering and Measurement (ESEM '24), October 24--25, 2024, Barcelona, Spain}
\acmDOI{10.1145/3674805.3695403}
\acmISBN{979-8-4007-1047-6/24/10}

\usepackage{multirow}
\usepackage{makecell}
\usepackage{subfigure}
\usepackage{balance}
\usepackage{listings}
\usepackage{color, xcolor}
\usepackage{algorithm2e}
\usepackage{framed}
\AtBeginDocument{%
	\providecommand\BibTeX{{%
			\normalfont B\kern-0.5em{\scshape i\kern-0.25em b}\kern-0.8em\TeX}}}

\begin{document}
	\begin{sloppypar}
		
		\title[Reducing Events to Augment Log-based Anomaly Detection Models: An Empirical Study]{Reducing Events to Augment Log-based Anomaly \\ Detection Models: An Empirical Study}
		
		\author{Lingzhe Zhang}
		\affiliation{%
			\institution{Peking University}
			\city{Beijing}
			\country{China}}
		\orcid{0009-0005-9500-4489}
		\email{zhang.lingzhe@stu.pku.edu.cn}
		
		\author{Tong Jia$^{\ast}$}
		\thanks{*Corresponding author}
		\affiliation{%
			\institution{Peking University}
			\city{Beijing}
			\country{China}}
		\orcid{0000-0002-5946-9829}
		\email{jia.tong@pku.edu.cn}
		
		\author{Kangjin Wang}
		\affiliation{%
			\institution{Alibaba Group}
			\city{Beijing}
			\country{China}}
		\email{kangjin.wkj@alibaba-inc.com}
		
		\author{Mengxi Jia}
		\affiliation{%
			\institution{Peking University}
			\city{Beijing}
			\country{China}}
		\orcid{0000-0002-0979-9803}
		\email{mxjia@pku.edu.cn}
		
		\author{Yong Yang}
		\affiliation{%
			\institution{Peking University}
			\city{Beijing}
			\country{China}}
		\orcid{0000-0001-9667-2423}
		\email{yang.yong@pku.edu.cn}
		
		\author{Ying Li$^{\ast}$}
		\affiliation{%
			\institution{Peking University}
			\city{Beijing}
			\country{China}}
		\orcid{0000-0002-6278-2357}
		\email{li.ying@pku.edu.cn}
		
		\renewcommand{\shortauthors}{Lingzhe Zhang et al.}
		
		\begin{abstract}
			As software systems grow increasingly intricate, the precise detection of anomalies have become both essential and challenging. Current log-based anomaly detection methods depend heavily on vast amounts of log data leading to inefficient inference and potential misguidance by noise logs. However, the quantitative effects of log reduction on the effectiveness of anomaly detection remain unexplored. Therefore, we first conduct a comprehensive study on six distinct models spanning three datasets. Through the study, the impact of log quantity and their effectiveness in representing anomalies is qualifies, uncovering three distinctive log event types that differently influence model performance. Drawing from these insights, we propose LogCleaner: an efficient methodology for the automatic reduction of log events in the context of anomaly detection. Serving as middleware between software systems and models, LogCleaner continuously updates and filters \textit{anti-events} and \textit{duplicative-events} in the raw generated logs. Experimental outcomes highlight LogCleaner's capability to reduce over 70\% of log events in anomaly detection, accelerating the model's inference speed by approximately 300\%, and universally improving the performance of models for anomaly detection.
		\end{abstract}

		
		
		\maketitle
		
		\section{Lay Abstract}
		
		As software systems become more complex, detecting problems or unusual behaviors (called anomalies) in these systems is both critical and difficult. Most current methods for finding these anomalies rely on processing huge amounts of log data, which can slow down the process and may lead to inaccurate results due to unnecessary or noisy logs. Despite this, the impact of reducing log data on anomaly detection hasn’t been studied much.
		
		To address this gap, we conduct a detailed study using six different models and three datasets to see how the quantity and quality of log data affect the ability to detect anomalies. We discover that not all log events are equally important—some help models perform better, while others have little or even negative effects. Based on these findings, we develop a tool called LogCleaner.
		
		LogCleaner is designed to automatically reduce the number of log events while still maintaining the important information needed for detecting anomalies. It works as a middle layer between the software system and the detection models, continuously removing unnecessary events (which we call "anti-events" and "duplicative-events") from the raw logs.
		
		Our experiments show that LogCleaner can remove more than 70\% of log events without hurting the ability to detect anomalies. In fact, by reducing the noise, it speeds up the models by about 300\% and improves their overall performance. LogCleaner offers a practical solution for developers and engineers looking to make anomaly detection faster and more accurate.
		
		\section{Introduction}
		
		Modern software systems are becoming increasingly complex, leading to more frequent failures that can cause considerable losses even during short periods of unavailability\cite{elliot2014devops, zhang2024survey}. Detecting anomalies accurately has therefore become critical for ensuring reliable and continuously available services. System logs provide valuable runtime information about software states and events, making them an indispensable resource for log-based anomaly detection approaches. With the ability to pinpoint failures and prevent further deterioration, log-based anomaly detection have garnered significant attention as important ways to maintain highly secure and resilient software systems in the face of rising complexity.
		
		In recent years, anomaly detection based on system logs has gained significant research attention. These log-based anomaly detection models can be broadly classified into two categories: supervised models\cite{zhang2019robust, yang2021semi, lu2018detecting, zhang2024multivariate} and unsupervised models\cite{babenko2009ava, du2017deeplog, yin2020improving, jia2021logflash,guo2021logbert, jia2022augmenting, kim2020automatic, meng2019loganomaly, landauer2023deep}. Supervised models, such as RobustLog\cite{zhang2019robust}, necessitate labeled data comprising both normal and abnormal instances to construct their predictive frameworks. In contrast, unsupervised models detect deviations relying solely on standard data. They are primarily split into deep neural network-based\cite{du2017deeplog, kim2020automatic, meng2019loganomaly, yin2020improving} and graph-based models\cite{babenko2009ava, jia2021logflash,guo2021logbert, jia2022augmenting, kim2020automatic}.
		
		Despite the promising results demonstrated by these anomaly detection methods, they directly leverage extensive log data generated by software systems, leading to the following practical challenges:
		
		\begin{itemize}
			\item \textbf{Inefficient Inference:} With an increasing number of logs, the model's inference speed tends to slow down. If a substantial portion of these original logs consists of irrelevant entries, it can result in unnecessary degradation of inference speed and resource wastage. \cite{yu2023logreducer}.
			\item \textbf{Misleading by Noise Logs:} It is acknowledged that having more logs provides a wealth of information, but in reality, many logs are of low quality, and some even contain noise. This can mislead the model \cite{zhang2019robust, li2022swisslog}.
		\end{itemize}
		
		In fact, not all logs generated by software systems are essential. However, the quantitative effects of log reduction on the effectiveness of anomaly detection remain unexplored. The significance of various log types and the subsequent performance trade-offs post their elimination remain uncertain.
		
		To fill this significant gap, we conduct an empirical study to quantify the impact of log reduction on anomaly detection The investigation spans six anomaly detection models (LR, SVM, Decision Tree, Isolation Forest, RobustLog\cite{zhang2019robust}, PLELog\cite{yang2021semi}) applied to three datasets\cite{he2023loghub} (HDFS, BGL, Thunderbird). We design two approaches: a retry-based method and a clustering-based approach, to validate the extent of log event reduction possible under constrained model performance degradation thresholds. The results reveal that for anomaly detection models, log events can be significantly reduced, and the reduction of logs can even enhance model effectiveness. In extreme cases, such as the Thunderbird dataset, a single log event (originally 1406 log events) can identify most anomalies.
		
		Furthermore, this work conducts an in-depth analysis of reducible log events for anomaly detection. The events are categorized into \textit{anti-events} and \textit{duplicative-events} based on whether their removal improves or does not affect model performance. Additionally, whose removal degrades model effectiveness are identifies as \textit{key-events}.
		
		Building on the findings of the empirical study, we introduce \textbf{LogCleaner}, a comprehensive methodology designed for the automatic reduction and reporting of \textit{anti-events} and \textit{duplicative-events} in log events, specifically tailored for anomaly detection. LogCleaner is divided into an profiling and an online component. In the profiling part, it utilizes historical logs, applying TF-IDF to eliminate sporadic log events, then using mutual information to reduce \textit{anti-events}. Finally, it employs a graph-based clustering approach to eliminate \textit{duplicative-events}, resulting in a reduced event set. In the online part, it functions as middleware between software systems and models, streamlining raw generated logs to reduced logs using the reduced event set. The reduced logs are then employed for anomaly detection. Additionally, whenever there is a variation in the code of the software system, the system's logs and existing labels are re-extracted for re-profiling. This re-profiling process enables LogCleaner to adapt effectively to potential future anomalies.
		
		We evaluate LogCleaner's effectiveness across the aforementioned models and datasets. Results show that LogCleaner can reduce over 70\% of log events in anomaly detection, accelerate the model's inference speed by approximately 300\%, while universally improve the performance of models for both anomaly detection. In summary, the contributions are as follows:
		
		\begin{itemize}
			\item We conduct a comprehensive study to quantify the impact of log event reduction on anomaly detection model effectiveness. Our findings reveal the remarkable extent to which the number of log events can be reduced without compromising model performance. Furthermore, our empirical study categorizes log events into \textit{key-events}, \textit{anti-events}, and \textit{duplicative-events} based on the impact of their removal on model performance.
			\item Inspired by the findings, we introduce \textbf{LogCleaner}, an efficient methodology for the automatic reduction of log events in the context of anomaly detection. Serving as middleware between software systems and models, LogCleaner continuously updates and filters \textit{anti-events} and \textit{duplicative-events} in the raw generated logs.
			\item We validate LogCleaner's effectiveness across 6 anomaly detection models on 3 datasets. Experiments demonstrate LogCleaner universally improves detection model performance while reducing over 70\% of log events and accelerating the model's inference speed by approximately 300\%.
		\end{itemize}
		
		\section{Background}
		
		This section provides background on log-based anomaly detection, introduces the models that will be used in the empirical study, and outlines the common overall framework for log-based anomaly detection.
		
		\subsection{Anomaly Detection} \label{anomaly-detection}
		
		\begin{figure*}[htbp]
			\centering
			\includegraphics[width=\textwidth]{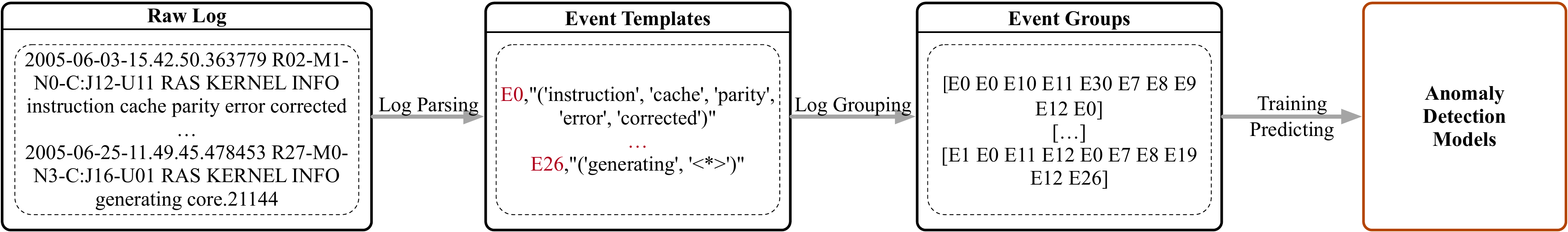}
			\caption{Log-based Anomaly Detection: The Common Workflow}
			\label{fig: common-workflow}
		\end{figure*}
		
		Anomaly detection\cite{zhang2019robust, yang2021semi, lu2018detecting, babenko2009ava, du2017deeplog, yin2020improving, jia2021logflash,guo2021logbert, jia2022augmenting} aims to identify irregularities in system behavior. Detection methods are broadly categorized into supervised and unsupervised models. Supervised models\cite{zhang2019robust, yang2021semi, lu2018detecting}, like \textbf{RobustLog}\cite{zhang2019robust}, require labeled data that includes both normal and abnormal examples to form predictive frameworks. \textbf{PLELog}\cite{yang2021semi} addresses the issue of insufficient labels via probabilistic label estimation and designs an attention-based GRU neural network to detect anomalies. Loglizer\cite{he2016experience} offers a comprehensive toolkit featuring several machine-learning-based log analysis models designed for automated anomaly detection, including linear regression (\textbf{LR}), \textbf{SVM}, \textbf{Decision Tree}, \textbf{Isolation Forest}. In contrast, unsupervised models\cite{du2017deeplog, yin2020improving, jia2021logflash,guo2021logbert, jia2022augmenting} identify deviations based only on standard data. In this paper, we focus on supervised models in analysis, given that labeled data allows a systematic assessment of log event reduction's impact on model performance. Unsupervised approaches cannot conclusively link performance changes to specific log event removals due to the absence of labels.
		
		\subsection{The Common Workflow}
		
		Despite that the target and approaches of anomaly detection are quite different, they share share common workflow\cite{jia2022augmenting, le2022log}. As shown in figure~\ref{fig: common-workflow}, the framework consists of three steps: (1) log parsing, (2) log grouping, (3) anomaly detection. 
		
		\subsubsection{Log parsing} Raw logs consist of semi-structured text encompassing various fields like timestamps and severity levels. For the benefit of downstream tasks, log parsing is employed to transform each log message into a distinct event template, which includes a constant part paired with variable parameters. For example, the log template "E0,('instruction', 'cache', 'parity', 'error', 'corrected')" can be extracted from the log message “2005-06-03-15.42.50.363779 R02-M1-N0-C:J12-U11 RAS KERNEL INFO instruction cache parity error corrected" in figure~\ref{fig: common-workflow}. There are many log parsing methods, based on frequent pattern mining\cite{dai2020logram, nagappan2010abstracting, sedki2022effective, yu2023brain}, clustering\cite{hamooni2016logmine, shima2016length, tang2011logsig, nedelkoski2021self}, and heuristics\cite{he2017drain, jiang2008abstracting, makanju2009clustering}. This paper utilizes the \textbf{Brain}\cite{yu2023brain} implemented by Logparser\cite{zhu2019tools,  he2016evaluation}.
		
		\subsubsection{Log grouping} After being parsed into event templates, log data can be organized into sequence groups using session, sliding, or fixed windows. Determining an optimal window size is challenging. For instance, a small window size might impede the models' ability to recognize anomalies that stretch across multiple sequences. Conversely, a large window size could lead to log sequences encompassing multiple anomalies, thereby complicating the detection process\cite{le2022log}. This study adopts both session-based and fixed windows of 100 logs, aligning parameters with those presented in the survey\cite{le2022log}.
		
		\subsubsection{Anomaly detection} After converting log events into sequences, they are processed by the previously mentioned anomaly detection models. These models undergo profiling training and then facilitate online prediction.
		
		\section{Study Design}
		
		This section outlines the datasets and models under evaluation and provides an overview of the methodology adopted for the empirical study.
		
		\subsection{Datasets}
		
		In our assessment of models for log-based anomaly detection, three datasets are employed\cite{he2023loghub}: HDFS, BGL and Thunderbird. The details of each dataset are as follows:
		
		\textbf{HDFS} dataset originates from over 200 Amazon EC2 nodes. It encompasses a total of 11,175,629 log messages. These messages are grouped into distinct log windows based on their block\_id, representing individual program executions within the HDFS system. Notably, 16,838 log blocks (amounting to 2.93\%) within this dataset signify system anomalies.
		
		\textbf{BGL} dataset is derived from a supercomputing system and was gathered by Lawrence Livermore National Labs (LLNL). It comprises a total of 4,747,963 log messages. Every message within the BGL dataset has been manually categorized as either normal or anomalous. Notably, of these, 348,460 log messages (representing 7.34\%) are marked as anomalous.
		
		\textbf{Thunderbird} dataset is an open collection of logs sourced from the Thunderbird supercomputer at Sandia National Labs (SNL). This dataset encompasses both regular and anomalous messages, each of which has been manually classified. While the Thunderbird dataset encompasses a massive collection of over 200 million log messages, this paper opts to use an initial continuous subset of 10 million log lines for the sake of computational efficiency. Notably, this subset includes 353,794 anomalous log messages, constituting 3.53\% of the total.
		
		\subsection{Evaluated Models}
		
		In this study, we evaluate the six representative models described in section \ref{anomaly-detection}. The source code for all anomaly detection models\cite{he2016experience, github:logdeep, yang2021semi} are public. In terms of log parsing, we employ the Brain\cite{yu2023brain} method as implemented by Logparser\cite{zhu2019tools, he2016evaluation}. For log grouping, we adopt different strategies based on the dataset: session-based windows are applied to the HDFS dataset, while for the BGL and Thunderbird dataset, we utilize fixed windows comprising 100 logs.
		
		\subsection{Approach}
		
		As previously mentioned, the aim of study is to quantify the effect of log event reduction on the effectiveness of anomaly detection models. To achieve this, we introduce two empirical study methodologies: the Retry-based approach and the Cluster-based approach.
		
		\begin{figure}[htbp]
			\centering
			\includegraphics[width=1\linewidth]{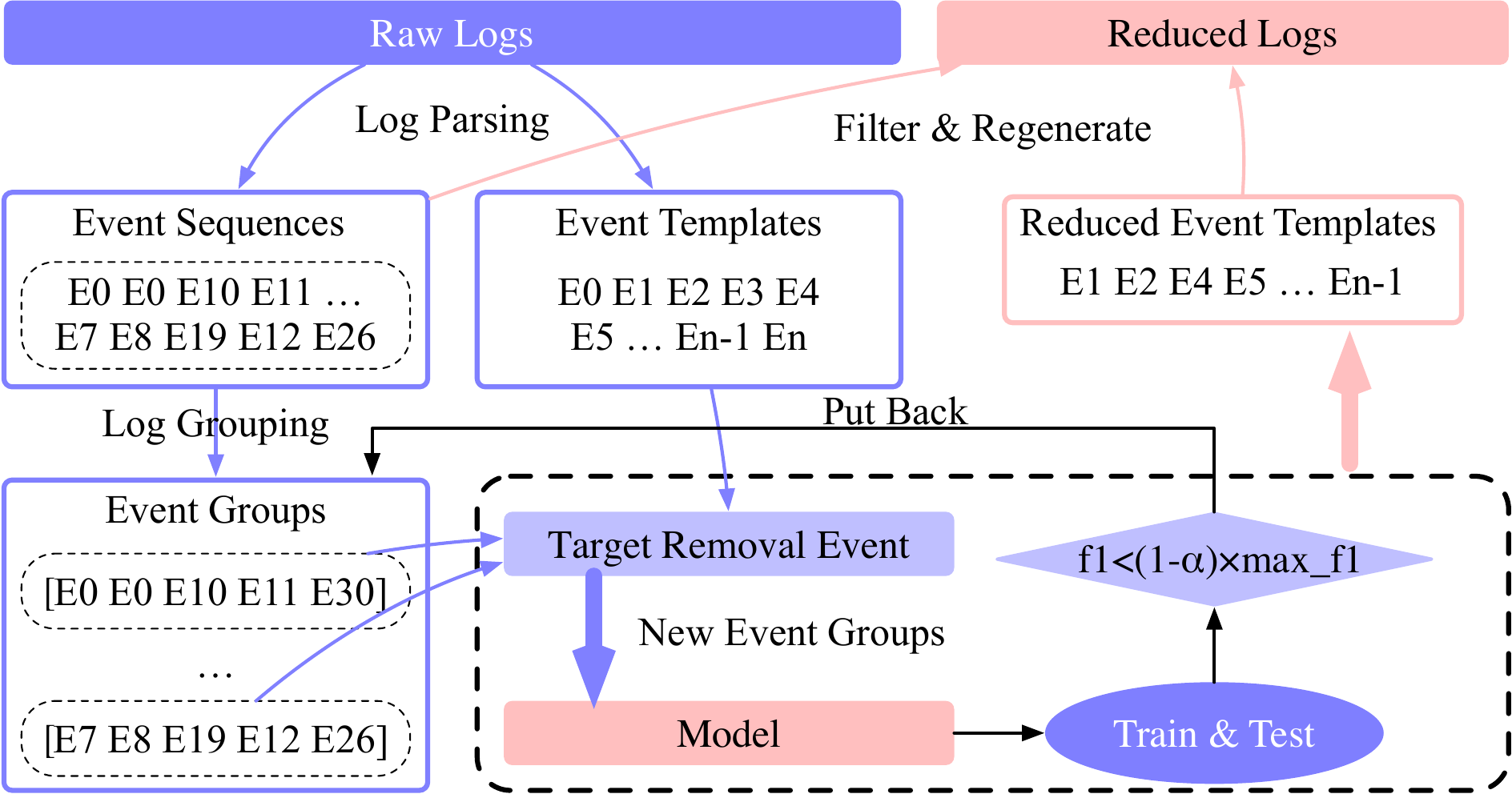}
			\caption{Process of Retry-based Approach}
			\label{fig:retry-based}
		\end{figure}
		
		\subsubsection{Retry-based approach}
		
		The core concept behind the Retry-based approach is to iteratively remove one log event at a time and then retrain the model to assess its effectiveness. If the model's effectiveness decreases after the removal, that particular log event is retained; otherwise, it's deemed useless.
		
		As illustrated in figure~\ref{fig:retry-based}, this approach initially employs the previously mentioned log parsing tool to extract event sequences and templates. Following this, the respective log grouping algorithm is utilized to organize these events into groups. Sequentially, it attempts to eliminate one log event from the event templates (designated as the target removal event), and simultaneously remove the corresponding event from each event group. It's crucial to note that after removing an event, the log grouping algorithm isn't re-executed. This ensures the preservation of labels for each event group. Thus, even if an event group is devoid of any events, its associated label is still retained.
		
		\begin{equation}
			f1 < (1-\alpha) \times f1_{max}
			\label{eq: f1-degrade}
		\end{equation}
		
		The regenerated event groups are subsequently fed into the respective model for retraining and testing. From this, it obtains metrics such as precision, recall, and the F1-score. If the model's performance meets the criteria defined in equation~\ref{eq: f1-degrade}, it indicates that the removal of that particular event impacts the model's effectiveness. As a result, this event is reintegrated into both the event templates and event groups. On the other hand, if the event doesn't significantly influence the performance, it is deemed redundant and removed, with the F1-score at that point recorded as $f1_{max}$. Here, $\alpha$ represents the permissible threshold for performance degradation. It's important to highlight that minor temporary performance dips during the model's training and testing phase don't necessarily signify a permanent degradation in model efficacy. Such deviations might merely be due to natural random fluctuations. Thus, even though it establishes a threshold with $\alpha$, the overall model effectiveness might not necessarily decline upon the completion of experiments.
		
		\subsubsection{Clustering-based approach}
		
		The Retry-based approach can produce near-optimal results. However, its necessity to retrain the model every time an event is removed becomes prohibitively time-consuming when dealing with datasets that have numerous events and a large volume of log events. This is because most model training durations are directly proportional to the volume of log events. For instance, considering the Thunderbird dataset, which consists of 1,406 event templates, if the SVM model initially takes around 10 minutes for each train-test iteration, completing the entire experiment will demand almost 10 days of computational time. Moreover, it's essential to highlight that many experiments require multiple runs to ensure consistent and reliable results.
		
\begin{figure}[htbp]
			\centering
			\includegraphics[width=1\linewidth]{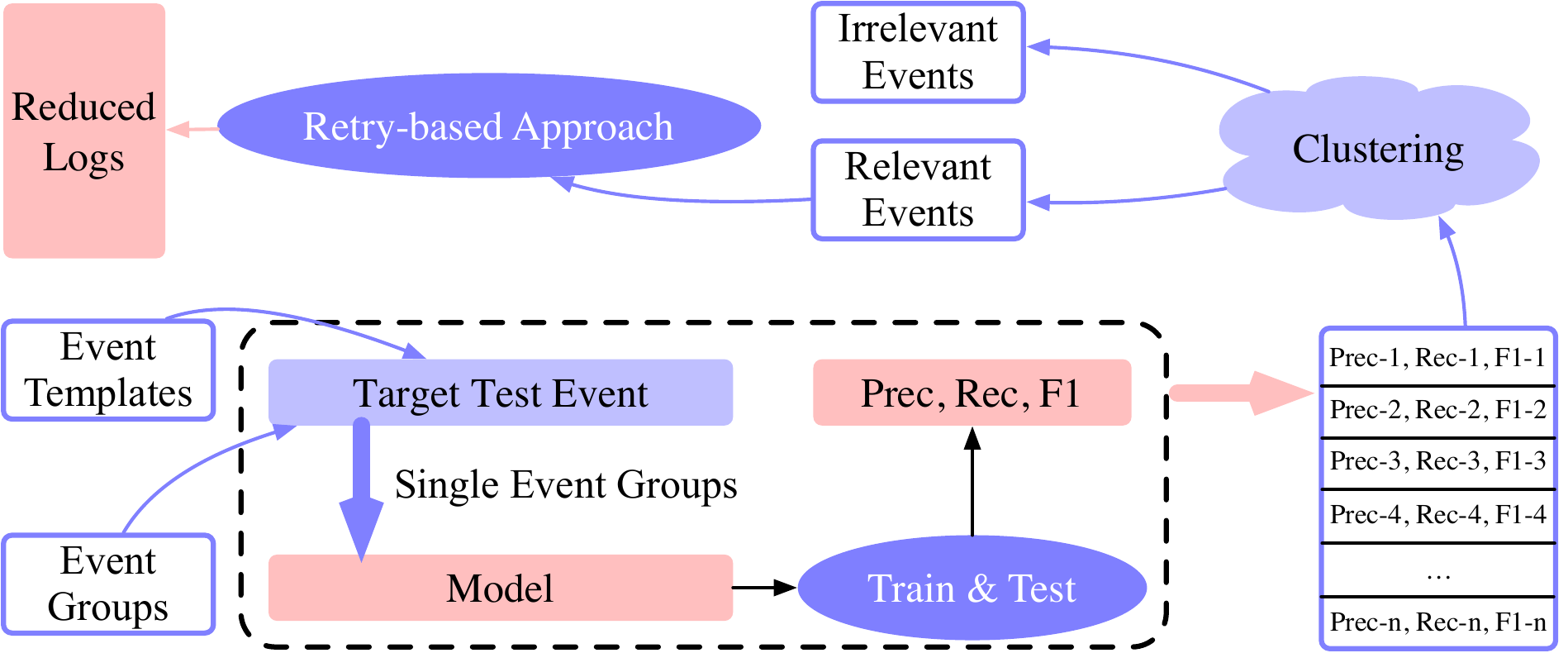}
			\caption{Process of Clustering-based Approach}
			\label{fig:cluster-based}
		\end{figure}

		Thus, to expedite the categorization and filtering of irrelevant log events, we introduce the Clustering-based approach. As illustrated in figure~\ref{fig:cluster-based}, it begins by extracting all log templates. Sequentially, it identifies each log event within the event templates as the Target Test Event. For each event group, only the corresponding Target Test Event is retained. These single event groups are then subjected to the specific model for retraining and testing. The resulting precision, recall, and F1-score are documented for every iteration. Ultimately, it obtains the precision, recall, and F1-score associated with each individual log event. Using a clustering algorithm (KMeans in this paper), based on the precision, recall, and F1-score, it classifies log events into two categories: irrelevant and relevant events. Finally, within the scope of relevant events, the Retry-based approach is executed.
		
		In summary, although the Retry-based approach can yield near-optimal results, it runs slowly when there are many event templates (as the model needs to be retrained every time an event is removed). Consequently, in our subsequent experiments, we employ the Retry-based approach for the HDFS dataset. For the BGL and Thunderbird datasets, we opt for the Clustering-based approach.
		
		\subsection{Research Questions} \label{rq}
		
		The objective of this research is to quantify the impact of log event reduction on the performance of anomaly detection models. Several research questions are formulated to guide the investigation, leveraging the experimental approaches previously described.
		
		\textbf{RQ1: To what extent can each anomaly detection method reduce log events?} We aim to assess the extent to which each method can reduce log events while maintaining the performance of existing anomaly detection approaches. For this purpose, useless otherwise mentioned, we conduct experiments on the studied models across various datasets with $\alpha=0.02$. This parameter choice ensures that the natural random fluctuations do not lead to false identification of an event as relevant, while also ensuring that the model's performance does not degrade significantly.
		
		\textbf{RQ2: How does log event reduction impact the performance of existing anomaly detection approaches?} In RQ1, we conduct experiments using $\alpha=0.02$, indicating a slight decline in the model's performance. However, as previously analyzed, this isn't necessarily the case. Hence, in this research question, we will further delve into a quantitative analysis of the impact of event reduction on the performance of anomaly detection approaches.
		
		\textbf{RQ3: What types of log events can be reduced without degrading anomaly detection performance?} In RQ1 and RQ2, we validate the quantity of log events that can be reduced and provide the performance of the model post-reduction. However, we also wish to analyze how log events should be distinguished in scenarios without access to the source code. Therefore, in this research question, we conduct several case studies on anomaly detection models to verify the different types of log events.
		
		\section{Empirical Results}
		
		This section presents and addresses the research questions proposed in section~\ref{rq}.
		
		\subsection{RQ1: Reduction Extent of Log Events for Anomaly Detection Methods}
		
		For RQ1, we conduct experiments on the studied models across various datasets with $\alpha=0.02$, to investigate the potential reduction in the volume of log events and lines. The experimental results are shown in table~\ref{tab: volumn-reduction-detection}.
		
		\begin{table}[htbp]
			\centering
			\caption{Data Volume Reduction in Anomaly Detection Across Various Models and Datasets ($\alpha=0.02$)}
			\label{tab: volumn-reduction-detection}
			\begin{tabular}{ccccc}
				\toprule
				\multicolumn{2}{c}{\textbf{Model}} & HDFS & BGL  & Thunderbird \\
				\midrule
				\multirow{2}*{LR} & \textit{events} & 55.17\% & 97.53\% & 99.93\%\\
				~ & \textit{lines} & 84.37\% & 98.94\% & 96.45\%\\
				\midrule
				\multirow{2}*{SVM} & \textit{events} & 65.52\% & 95.78\% & 99.93\%\\
				~ & \textit{lines} & 84.58\% & 98.90\% & 96.45\%\\
				\midrule
				\multirow{2}*{Decision Tree} & \textit{events} & 75.86\% & 92.44\% & 99.57\%\\
				~ & \textit{lines} & 72.09\% & 91.96\% & 94.04\%\\
				\midrule
				\multirow{2}*{Isolation Forest} & \textit{events} & 68.97\% & 73.55\% & 93.60\%\\
				~ & \textit{lines} & 76.20\% & 85.44\% & 85.38\%\\
				\midrule
				\multirow{2}*{RobustLog} & \textit{events} & 58.62\% & 94.48\% & 99.93\%\\
				~ & \textit{lines} & 54.60\% & 50.81\% & 96.45\%\\
				\midrule
				\multirow{2}*{PLELog} & \textit{events} & 65.52\% & 94.62\% & 99.93\%\\
				~ & \textit{lines} & 68.86\% & 97.00\% & 96.45\%\\
				\bottomrule
			\end{tabular}
		\end{table}
		
		It can be find that, while maintaining consistent model performance, all anomaly detection models show a significant reduction in log events. The reduction ranges from a minimum of 55.17\% (with the LR model on the HDFS dataset) to more than 99\% (in the Thunderbird dataset). This suggests that a majority of log events in the HDFS, BGL, and Thunderbird datasets are, in fact, superfluous.
		
		We also explore the reduction of log lines (where one log event corresponds to multiple lines of actual printed logs, as each log event essentially corresponds to each line of code where developers write print statements). Even though the model LR in the HDFS dataset only reduced 55.17\% of log events, the actual reduction in log lines reached 84.37\%. This indicates that the eliminated log events are of high frequency, constituting a large proportion of the entire dataset. Globally, all anomaly detection models show a significant reduction in log lines. To further underscore the significance of these results, we take the SVM model on the BGL dataset as an example. The original BGL log file is 743.19MB in size. After reduction, it is whittled down to just 7.87MB. This not only dramatically accelerates the model's training speed but can also provide feedback to system developers, thereby reducing the overhead associated with log collection.
		
		\begin{table}[htbp]
			\centering
			\caption{Remaining Log Events After Reduction in Anomaly Detection Across Various Models and Datasets ($\alpha=0.02$)}
			\label{tab: remain-detection}
			\begin{tabular}{cccc}
				\toprule
				\multirow{2}*{\textbf{Model}} & HDFS & BGL  & Thunderbird \\
				~ & (29 events) & (688 events) & (1406 events) \\
				\midrule
				LR & 13 & 17 & 1\\
				SVM & 10 & 29 & 1\\
				Decision Tree & 7 & 52 & 6\\
				Isolation Forest & 9 & 182 & 90\\
				RobustLog & 12 & 38 & 1\\
				PLELog & 10 & 37 & 1\\
				\bottomrule
			\end{tabular}
		\end{table}
		
		We also observe a pervasive and startling reduction in some datasets. Therefore, as depicted in table~\ref{tab: remain-detection}, we further examine the remaining log events after reduction. It's evident that the reason why various models have a relatively low reduction ratio on the HDFS dataset is due to the dataset itself containing only 29 events. For the BGL dataset, apart from the Isolation Forest model, all other models require fewer than 50 out of the 688 events. As for the Thunderbird dataset, an even more remarkable result emerges: for models like LR, SVM, and RobustLog, they only require one out of the 1,406 events to achieve exceptionally high accuracy.
		
		\begin{table*}[t]
			\centering
			\caption{Comparison of Anomaly Detection Model Performance with/out event reduction ($\alpha=0.02$)}
			\label{tab:comparison-detection}
			\begin{tabular}{ccccccccccc}
				\toprule
				\multirow{2}*{\textbf{Model}} & & \multicolumn{3}{c}{HDFS} & \multicolumn{3}{c}{BGL}  & \multicolumn{3}{c}{Thunderbird} \\
				\\[-2ex]
				\cline{3-11}
				\\[-1.2ex]
				~ & ~ & \textbf{\textit{Precision}} & \textbf{\textit{Recall}} & \textbf{\textit{F1-Score}} & \textbf{\textit{Precision}} & \textbf{\textit{Recall}} & \textbf{\textit{F1-Score}} & \textbf{\textit{Precision}} & \textbf{\textit{Recall}} & \textbf{\textit{F1-Score}}\\
				\midrule
				\multirow{2}*{LR} & \textbf{w/o} & 0.952 & 0.711 & 0.814 & 0.193 & 0.829 & 0.313 & 0.971 & 0.995 & 0.983 \\
				~ & \textbf{w} & 0.948 & 0.970 & 0.959 & 0.982 & 0.456 & 0.623 & 1.000 & 0.999 & 0.999 \\
				\midrule
				\multirow{2}*{SVM} & \textbf{w/o} & 0.959 & 0.889 & 0.923 & 0.877 & 0.378 & 0.529 & 0.991 & 0.996 & 0.994 \\
				~ & \textbf{w} & 0.959 & 0.889 & 0.923 & 0.979 & 0.433 & 0.601 & 1.000 & 0.998 & 0.999 \\
				\midrule
				\multirow{2}*{Decision Tree} & \textbf{w/o} & 0.998 & 0.998 & 0.998 & 0.992 & 0.406 & 0.576 & 1.000 & 0.999 & 0.999 \\
				~ & \textbf{w} & 0.998 & 0.998 & 0.998 & 0.989 & 0.450 & 0.618 & 1.000 & 0.999 & 1.000 \\
				\midrule
				\multirow{2}*{Isolation Forest} & \textbf{w/o} & 0.822 & 0.742 & 0.780 & 0.613 & 0.166 & 0.261 & 0.005 & 0.001 & 0.002 \\
				~ & \textbf{w} & 0.936 & 0.911 & 0.923 & 0.917 & 0.240 & 0.381 & 0.775 & 0.097 & 0.173 \\
				\midrule
				\multirow{2}*{RobustLog} & \textbf{w/o} & 0.985 & 0.888 & 0.934 & 1.000 & 0.991 & 0.996 & 0.910 & 0.559 & 0.692 \\
				~ & \textbf{w} & 0.994 & 0.995 & 0.994 & 1.000 & 1.000 & 1.000 & 1.000 & 1.000 & 1.000 \\
				\midrule
				\multirow{2}*{PLELog} & \textbf{w/o} & 0.983 & 0.843 & 0.908 & 0.943 & 0.986 & 0.969 & 0.968 & 0.996 & 0.982 \\
				~ & \textbf{w} & 0.998 & 0.971 & 0.984 & 0.999 & 0.967 & 0.983 &1.000 & 0.996 & 0.998 \\
				\bottomrule
			\end{tabular}
		\end{table*}
		
		\begin{figure}[htbp]
			\centering
			\subfigure[Log Event Template]{
				\begin{minipage}{0.43\linewidth}
					\centering
					\includegraphics[width=\textwidth]{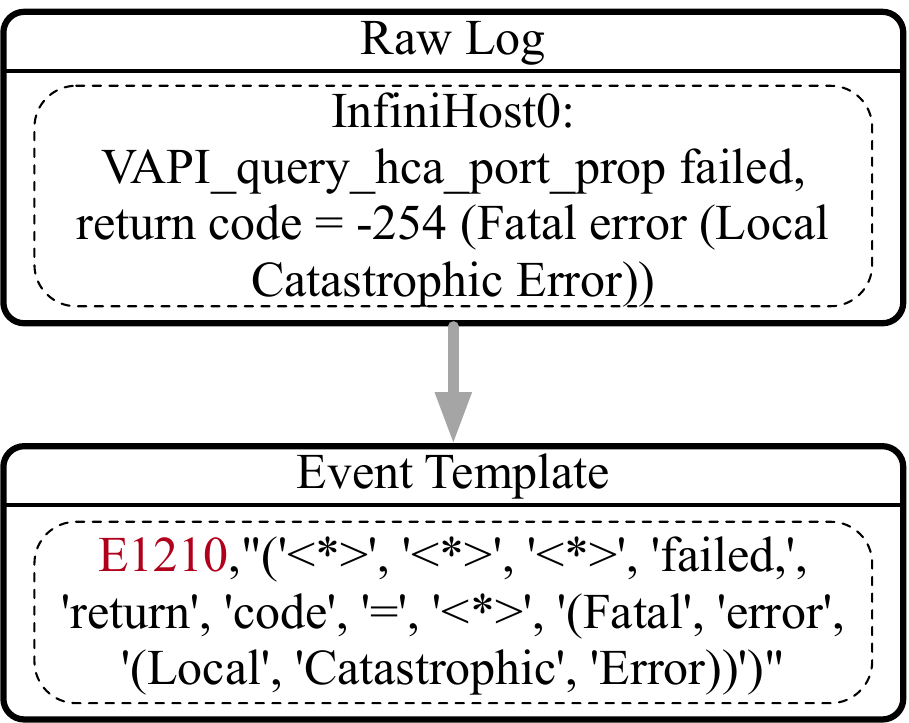}
					\label{fig: e1210-template}
				\end{minipage}
			}
			\subfigure[Performance]{
				\begin{minipage}{0.49\linewidth}
					\centering   
					\includegraphics[width=\textwidth]{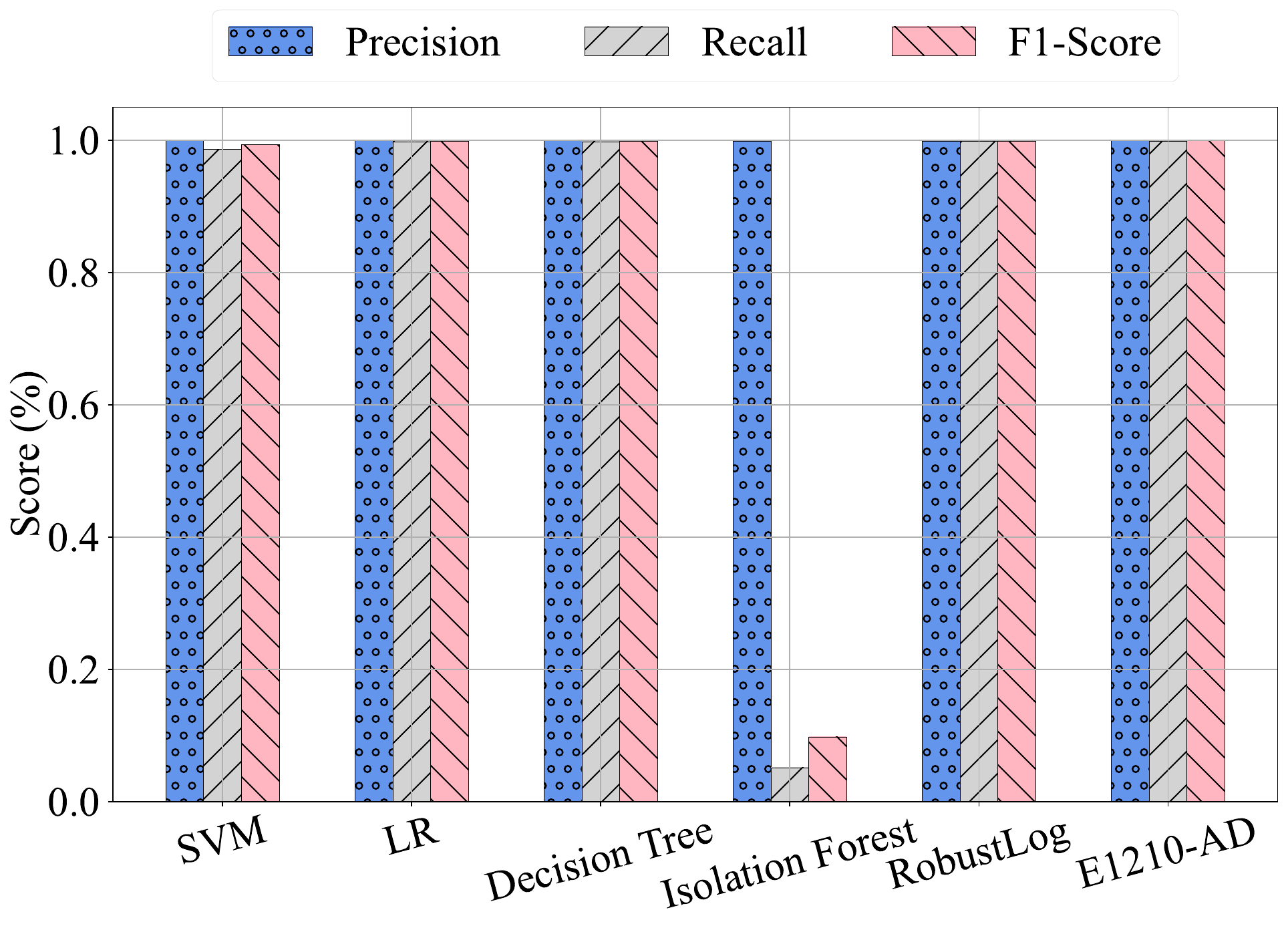}
					\label{fig: e1210-score}
				\end{minipage}
			}
			\caption{Analysis of the Only Remaining Event (E1210) and Its Impact on Model Performance}
			\label{fig: e1210}
		\end{figure}
		
		For the Thunderbird dataset, a notable observation is that the sole remaining event for these models is E1210, as illustrated in figure~\ref{fig: e1210-template}. We hypothesize that the occurrence of the E1210 log event could signify the presence of an anomaly within this dataset. To validate this, we specifically devise a heuristic method for detection, dubbed E1210-AD. The results confirm the speculation, as depicted in figure~\ref{fig: e1210-score}. Apart from the Isolation Forest model, the accuracy of other models essentially reaches above 98.5\%, aligning closely with the performance of E1210-AD.
		
		\begin{figure}[htbp]
			\centering
			\subfigure[HDFS]{
				\begin{minipage}{0.46\linewidth}
					\centering
					\includegraphics[width=\textwidth]{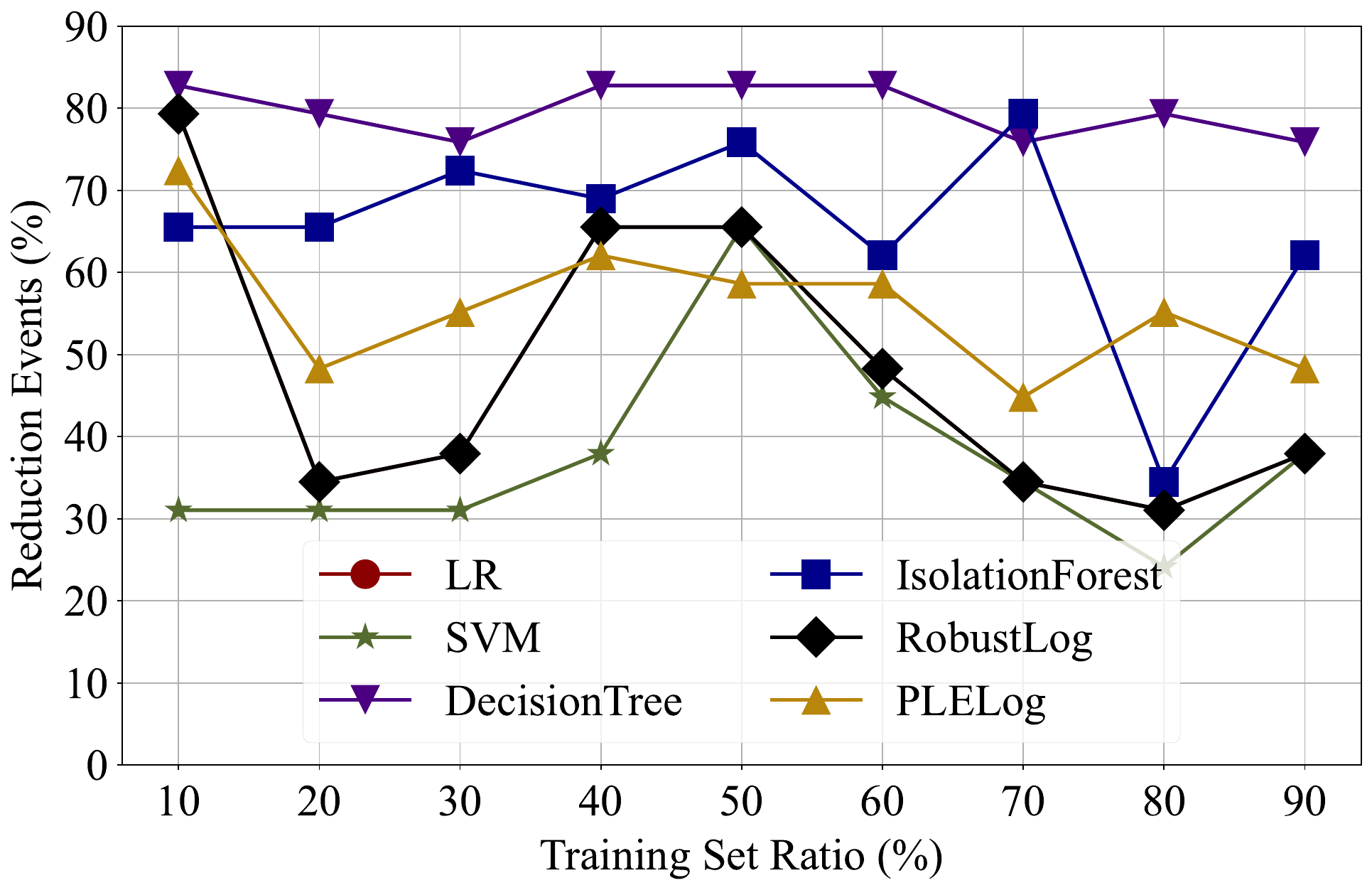}
					\label{fig: detection-update-ratio-hdfs}
				\end{minipage}
			}
			\subfigure[BGL]{
				\begin{minipage}{0.46\linewidth}
					\centering   
					\includegraphics[width=\textwidth]{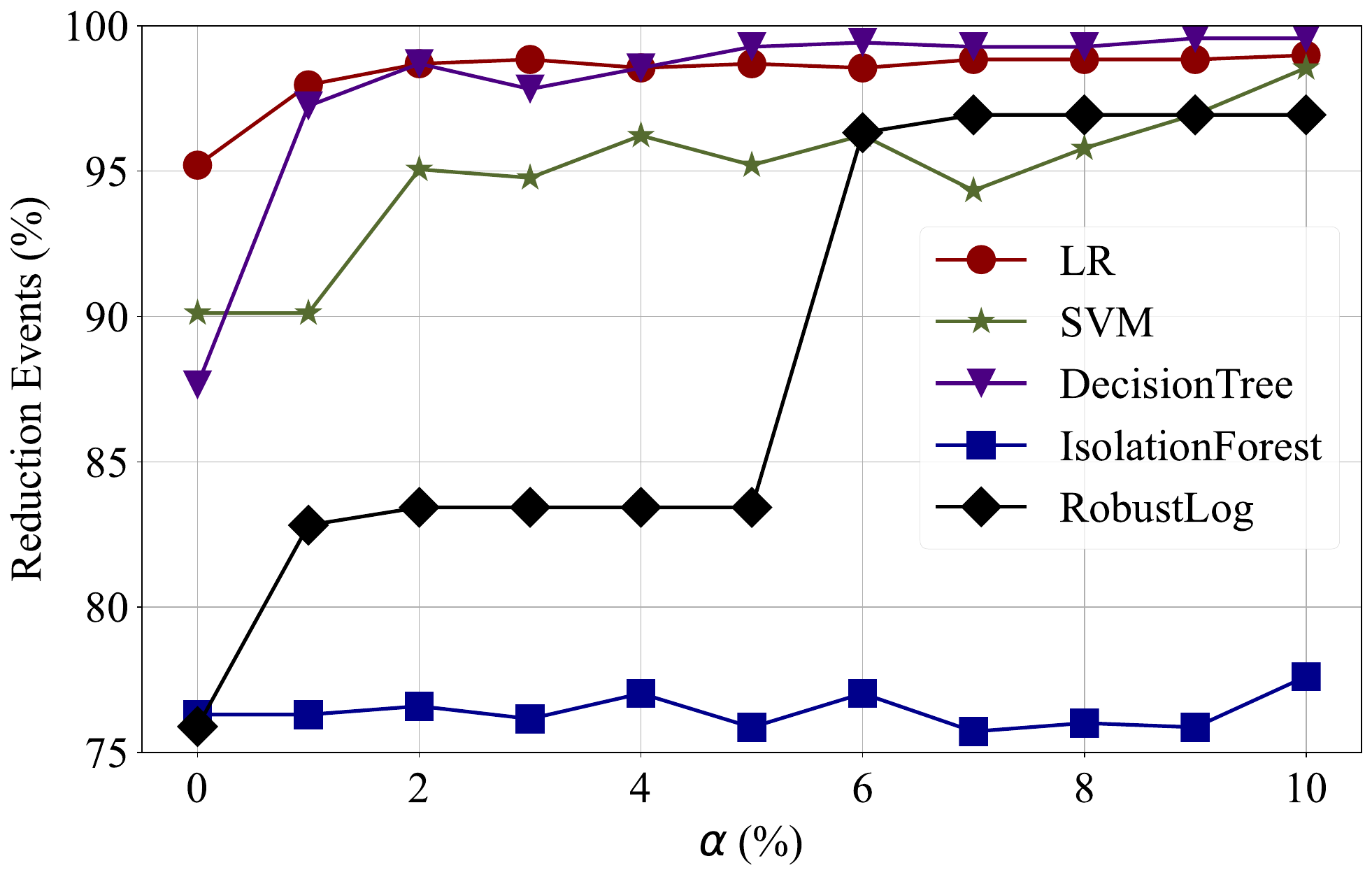}
					\label{fig: detection-update-ratio-bgl}
				\end{minipage}
			}
			\caption{Extent of Log Event Reduction in Anomaly Detection Methods Depending on the Variation of $\alpha$}
			\label{fig: detection-update-ratio}
		\end{figure}
		
		We further investigate how the quantity of reduced logs changes with varying values of $\alpha$. As shown in Figure~\ref{fig: detection-update-ratio}, for the HDFS dataset, the reduction remains stable when $\alpha \in {0, 0.01}$ and beyond, as lower values of $\alpha$ may be influenced by the model's random fluctuations, making it hard to assess log event relevance accurately.
		
		A similar trend is observed for the BGL dataset, with RobustLog showing a significant change at $\alpha = 0.05$, indicating that some events can be removed, but this may affect performance. In contrast, Isolation Forest remains consistent, likely due to its difficulty in detecting anomalies in the BGL dataset.
		
		\begin{framed}
			\noindent
			\textbf{Summary.} A large number of log events can be reduced in anomaly detection. In some extreme cases, it can even be reduced to a single event.
		\end{framed}
		
		\subsection{RQ2: Performance After Event Reduction}
		
		For RQ2, we conduct a detailed analysis of the performance of anomaly detection models with and without event reduction, setting $\alpha = 0.02$.
		
		As demonstrated in table~\ref{tab:comparison-detection}, the performance of nearly all models improved with event reduction. Some even experienced significant enhancements. For instance, RobustLog on the Thunderbird dataset initially has an F1-Score of 69.2\%. However, after event reduction, it soares to 100\%. For the Isolation Forest model, its original performance on the Thunderbird dataset is nearly negligible with an F1-Score of 0.02\%. However, after event reduction, this score improved to 17.3\%. Even more impressively, for the LR model on the BGL dataset, by balancing Precision and Recall (Precision shifted from 19.3\% to 98.2\%, and Recall shifted from 82.9\% to 45.6\%), the overall F1-Score nearly doubles compared to the initial performance.

		\begin{framed}
			\noindent
			\textbf{Summary.} For anomaly detection, after performing log reduction, the model's effectiveness can also be significantly improved.
		\end{framed}
		
		\subsection{RQ3: Types of Log Events That Can Be Reduced}
		
		In RQ1 and RQ2, we discover that, for anomaly detection, the model performance can improve to varying degrees with a significant reduction in log events. Thus, in RQ3, we delve deeper into this phenomenon by case study.
		
		Initially, we examine the reasons for the enhancement in model performance after event reduction. In a particular instance with the LR model in the HDFS dataset, upon removing the entry "\textit{E3,[*]Served block[*]to[*]}", there's a notable improvement: Precision by 0.63\%, Recall by 22.88\%, and F1-Score by 13.49\%. The reason for such a phenomenon is that this work discovers that this event appears in the normal label with a ratio of 23.97\% and in the abnormal label with a ratio of 21.54\%. This suggests that this particular log event acts as a distractor, potentially misleading the model's classification efforts. Furthermore, a similar pattern can be observed across all analyzed models.
		
		\begin{framed}
			\noindent
			\textbf{Finding 1.} In the dataset, there exists a type of event called \textbf{anti-event}. Its presence has no bearing on whether the system has generated an anomaly. Instead, it can mislead the model's classification.
		\end{framed}
		
		However, during experiments, it is observed that the number of anti-events is relatively small. In fact, the most frequently eliminated events belong to another category termed as duplicative-events. Taking the Decision Tree experiment on the HDFS dataset as an example: when using only \textit{E9}, it obtains a Precision of 100\%, Recall of 37.56\%, and F1-Score of 54.61\%. With only \textit{E11}, the Precision is 100\%, Recall is 37.55
		\%, and F1-Score is 54.59\%. When both \textit{E9} and \textit{E11} are used simultaneously, the metrics are Precision at 100\%, Recall at 37.56\%, and F1-Score at 54.61\%. These results suggest that in anomaly detection, certain log events can effectively substitute for others. In such cases, it can safely remove the redundant logs without any loss of information.
		
		\begin{framed}
			\noindent
			\textbf{Finding 2.} In anomaly detection, certain events can encompass the information of others. These overshadowed events can be safely removed without compromising model effectiveness. These events are termed as \textbf{duplicative-event}.
		\end{framed}

		\begin{figure*}[htbp]
			\centering
			\includegraphics[width=1\linewidth]{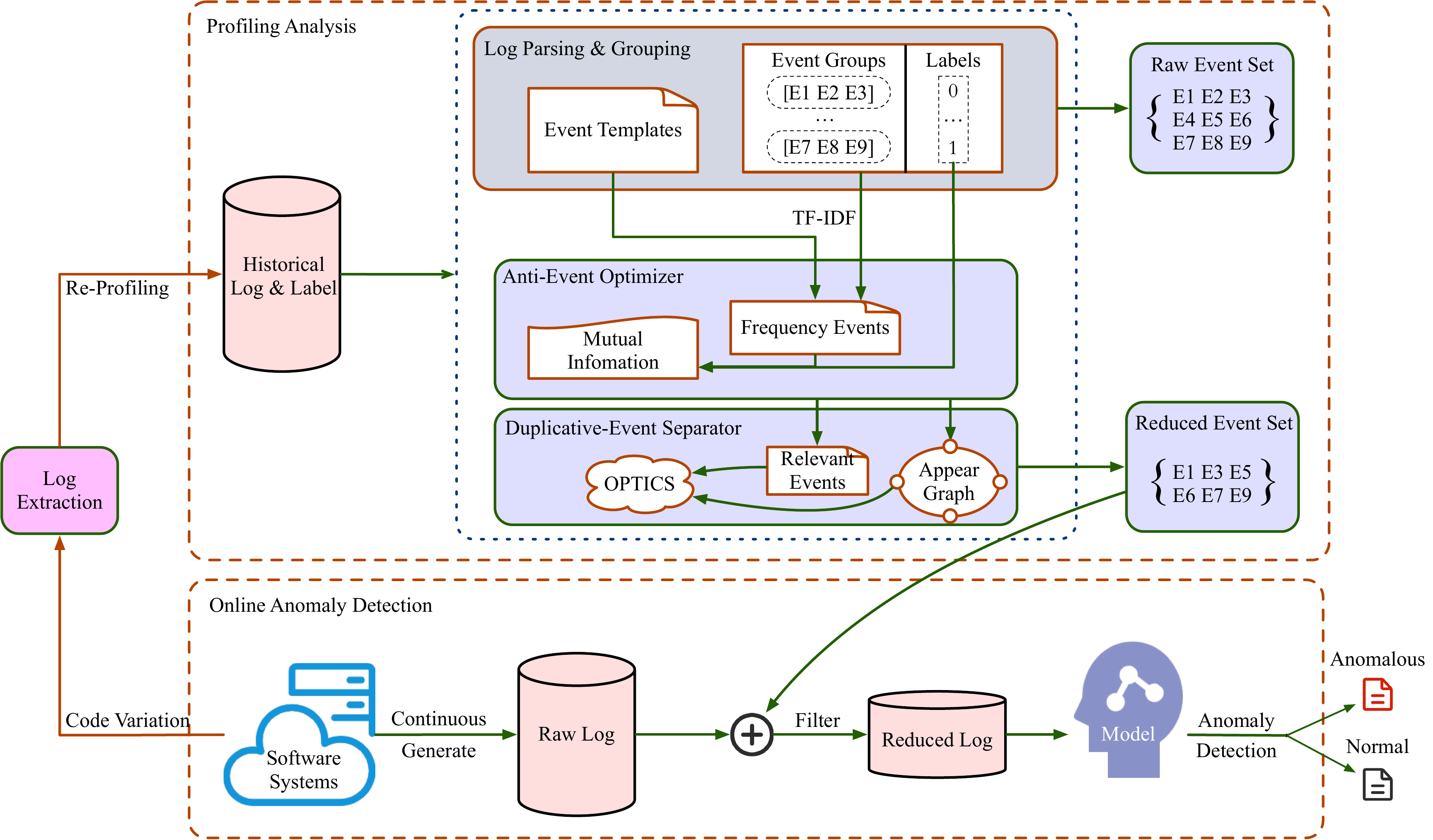}
			\caption{Workflow of LogCleaner}
			\label{fig: logcleaner}
		\end{figure*}
		
		Beyond the anti-events and duplicative-events that can be reduced, there remains a category of log events in the dataset that play a pivotal role in model effectiveness, termed as \textbf{key-event}. Taking the experiment with the Decision Tree on the HDFS dataset as an example, it can be found that when using \textit{E20} alone, the Precision is 95.85\%, Recall is 29.92\%, and F1-Score is 45.61\%. When using \textit{E26} alone, the Precision is 97.26\%, Recall is 59.46\%, and F1-Score is 73.80\%. However, when both \textit{E20} and \textit{E26} are used together, the Precision is 96.74\%, Recall is 88.24\%, and F1-Score soar to 92.30\%. This suggests that the system information reflected by E20 and E26 is complementary to each other. Such events are the ones that truly need to retain.
		
		\begin{framed}
			\noindent
			\textbf{Finding 3.} There exists a category of log events in the dataset that are crucial for model performance, with their information complementing each other. These are termed as \textbf{key-events}. It is these events that truly need to retain.
		\end{framed}
		
		\section{LogCleaner}
		
		Our empirical study identifies three types of log events that have different effects on the models. However, the experiments presented earlier required continuous model execution to determine the log events that can be reduced. Moreover, these methods do not provide an opportunity to reintroduce eliminated log events, even though they may represent potential future anomalies. In this section, we introduce \textbf{LogCleaner}, an automated approach to reduce log events without relying on model execution. Additionally, it allows for the reintroduction of some reduced log events when the system encounters false negatives, providing the model with the minimum log event set for current-state detection.
		
		As demonstrated in figure~\ref{fig: logcleaner},  LogCleaner is divided into an profiling and an online component. In the profiling phase, it aims to automatically generate a reduced event set. To achieve this, raw logs are initially parsed into event templates and grouped into event groups with corresponding labels. Subsequently, TF-IDF is applied to filter out infrequently occurring events. The remaining events (Frequency Events) undergo processing by the \textbf{Anti-Event Optimizer}, which utilizes both event groups and associated labels, employing mutual information to eliminate anti-events. The events surviving this stage (Relevant Events) then go through the \textbf{Duplicative-Event Separator}, where the OPTICS algorithm clusters similar events, retaining only one event within each cluster. Finally, the reduced event set are generated based on the retained events.
		
		In the online phase, LogCleaner serves as middleware between software systems and models, streamlining raw generated logs into reduced logs using the reduced event set generated in profiling phase. These reduced logs are then utilized for anomaly detection. Additionally, whenever there is a variation in the code of the software system, the system's logs and existing labels are re-extracted for re-profiling. This re-profiling process enables LogCleaner to adapt effectively to potential future anomalies.
		
		\subsection{Anti-Event Optimizer}
		
		In RQ3, it can be discovered that certain anti-events have no correlation with the occurrence of system anomalies or the specific anomalies that are triggered. Consequently, these anti-events negatively impact the model's classification performance.
		
		The analysis suggests that the presence or absence of such log events bears no relation to labels. Therefore, mutual information, a method from the feature selection domain, can be employed to estimate the relationship between each log event and its corresponding label.
		
		\begin{equation}
			\text{MI}(e;l) = \sum_{e,l}^{} p(e,l) \log \left( \frac{p(e,l)}{p(e)p(l)}\right)
			\label{eq: mi}
		\end{equation}
		
		As illustrated in equation~\ref{eq: mi}, all events are denoted as $e \in E$, all labels as $l \in L$, $p(e,l)$ represents the joint probability distribution of $E$ and $L$, while $p(e)$ and $p(l)$ are the marginal probability distributions of $E$ and $L$, respectively.
		
		Ultimately, as depicted in equation~\ref{eq: mi-sum}, the mutual information for each event $e$ is represented as the average of its mutual information with all labels $l$. Among them, events with $\text{MI}(e;L) \le \theta_{anti}$ are deemed as anti-events.
		
		\begin{equation}
			\text{MI}(e;L) = \frac{1}{|L|} \sum_{l \in L} \text{MI}(e;l)
			\label{eq: mi-sum}
		\end{equation}
		
		\subsection{Duplicative-Event Separator}
		
		For duplicative-events, LogCleaner initially constructs an appear graph based on the co-occurrence patterns of log events. This entails representing each log event as a vector ($e_i$). When two log events  ($e_i, e_j$) co-occur within a single event group, the weight of the edge ($w(e_i,e_j)$) between them in the graph is incremented.
		
		Next, LogCleaner employs the OPTICS algorithm, a density-based clustering method, to cluster the adjacency matrix of the aforementioned appear graph. Within OPTICS, it is believed that for a point to be considered as a core point, the number of log events in its neighborhood should satisfy $min\_samples \le \theta_{dup}$. For each cluster, LogCleaner retains the event with the highest 
		$\text{MI}(e;l)$ value as the representative event and remove the other events in the cluster. Furthermore, all outlier events are preserved as $E_{r}$. 
		
		\section{Experiment and Evaluation}
		
		This section evaluates the overall results, conduct an ablation study of LogCleaner, and assess the influence of hyperparameters.
		
		We perform experiments on the 6 models and 3 datasets previously discussed. Unless otherwise specified, LogCleaner utilizes TF-IDF to filter out events with a frequency below 0.1. The Anti-Event Optimizer's threshold, $\theta_{anti}$, is set to 0, while the threshold $\theta_{dup}$ for the Duplicative-Event Separator is set to 2.
		
		\subsection{Overall Evaluation Results}
		
		For anomaly detection, as illustrated in table~\ref{tab: volumn-reduction-detection-logcleaner}, the number of reducible events is significant across all datasets. For the Thunderbird and BGL datasets, the events are reduced by approximately 70\%. While this doesn't quite match the results from the previous empirical study, LogCleaner operates quickly and the reduced events are applicable across all models.
		
		\begingroup
		\setlength{\tabcolsep}{10pt}
		\begin{table}[htbp]
			\centering
			\caption{Data Volume Reduced by LogCleaner}
			\label{tab: volumn-reduction-detection-logcleaner}
			\begin{tabular}{c ccc}
				\toprule
				\multirow{2}{*}{\textbf{Type}} & \multicolumn{3}{c}{\textbf{Dataset}} \\
				\cmidrule(lr){2-4}
				& \textbf{HDFS} & \textbf{BGL} & \textbf{Thunderbird} \\
				\midrule
				events & 48.28\% & 73.13\% & 69.91\% \\
				lines  & 52.62\% & 24.66\% & 51.32\% \\
				\bottomrule
			\end{tabular}
		\end{table}
		\endgroup
		
		Furthermore, the performance of each model after applying LogCleaner is analyzed. As illustrated in Table~\ref{tab: upgrade-detection-logcleaner}, where LC\_RobustLog represents the RobustLog model enhanced with LogCleaner, and LC\_PLELog represents the PLELog model enhanced with LogCleaner. After applying LogCleaner, we observe significantly improved results compared to the original models. LC\_RobustLog achieves an increased F1-score of 5.17\%, 0.07\%, and 30.16\% on the HDFS, BGL, and Thunderbird datasets, respectively, in comparison to the RobustLog. Similarly, LC\_PLELog achieves an increased F1-score of 7.62\%, 1.41\%, and 1.61\% on the HDFS, BGL, and Thunderbird datasets, respectively, compared to the PLELog. The observed enhancements align with the findings from the previous empirical study.
		
		\begingroup
		\setlength{\tabcolsep}{8pt}
		\begin{table}[htbp]
			\centering
			\caption{Evaluation Results on Anomaly Detection Effectiveness}
			\label{tab: upgrade-detection-logcleaner}
			\begin{tabular}{ccccc}
				\toprule
				\textbf{Model} & & HDFS & BGL  & Thunderbird \\
				\midrule
				\multirow{3}*{Isolation Forest} 
				& P & 82.20\% & 61.30\% & 0.49\%\\
				~ & R & 74.20\% & 16.58\% & 0.10\%\\
				~ & F1 & 77.99\% & 26.10\% & 0.17\%\\
				\midrule
				\multirow{3}*{RobustLog} 
				& P & 98.50\% & 100.00\% & 90.96\%\\
				~ & R & 88.80\% & 99.14\% & 55.90\%\\
				~ & F1 & 93.40\% & 99.57\% & 69.25\%\\
				\midrule
				\multirow{3}*{PLELog} 
				& P & 98.34\% & 94.30\% & 96.84\%\\
				~ & R & 84.39\% & 99.82\% & 99.64\%\\
				~ & F1 & 90.83\% & 96.92\% & 98.22\%\\
				\midrule
				\multirow{3}*{LC\_RobustLog} 
				& P & 99.23\% & 100.00\% & 100.00\%\\
				~ & R & 97.91\% & 99.27\% & 98.63\%\\
				~ & F1 & 98.57\% & 99.64\% & 99.31\%\\
				\midrule
				\multirow{3}*{LC\_PLELog} 
				& P & 99.81\% & 99.98\% & 100.00\%\\
				~ & R & 97.13\% & 96.73\% & 99.62\%\\
				~ & F1 & 98.45\% & 98.33\% & 99.83\%\\
				\bottomrule
			\end{tabular}
		\end{table}
		\endgroup
		
		\subsection{Inference Time}
		
		\begingroup
		\setlength{\tabcolsep}{6pt}
		\begin{table}[htbp]
			\centering
			\caption{Inference Time Comparison with(out) LogCleaner in Anomaly Detection}
			\label{tab: inference-time}
			\begin{tabular}{ccccc}
				\toprule
				\textbf{Model} & & HDFS & BGL  & Thunderbird \\
				\midrule
				\multirow{2}*{LR} 
				& w/o & 3371.02 & 1052.37 & 11870.16\\
				~ & w & 1809.62 & 279.39 & 5313.95\\
				\midrule
				\multirow{2}*{SVM} 
				& w/o & 3150.26 & 1051.88 & 11869.10\\
				~ & w & 1906.75 & 280.33 & 5214.05\\
				\midrule
				\multirow{2}*{Decision Tree} 
				& w/o & 3666.74 & 1078.35 & 11870.16\\
				~ & w & 1443.89 & 264.69 & 5313.95\\
				\midrule
				\multirow{2}*{Isolation Forest} 
				& w/o & 12557.61 & 2343.57 & 32426.30\\
				~ & w & 10327.30 & 651.90 & 12161.74\\
				\midrule
				\multirow{2}*{RobustLog} 
				& w/o & 37571.02 & 10282.24 & 5859.79\\
				~ & w & 14428.25 & 2881.30 & 4131.11\\
				\midrule
				\multirow{2}*{PLELog} 
				& w/o & 10802.95 & 25306.99 & 15899.39\\
				~ & w & 7466.62 & 19303.02 & 9512.93\\
				\bottomrule
			\end{tabular}
		\end{table}
		\endgroup
		
		The substantial reduction in log events greatly enhances the model's inference speed. Therefore, we conduct experiments to validate the average inference time of the models before and after applying LogCleaner in the context of anomaly detection.
		
		As shown in Table~\ref{tab: inference-time}, we record the time each anomaly detection model takes to infer the entire test set across various datasets, measured in milliseconds. Clearly, the inference speed of all models significantly improved after applying LogCleaner, ranging from 21.59\% to 307.41\%.
		
		\subsection{Effectiveness of Event Reduction Components}
		
		\begingroup
		\setlength{\tabcolsep}{6pt}
		\begin{table}[htbp]
			\centering
			\caption{LogCleaner Ablation Experiment on Templates Reduction}
			\label{tab: ablation-logcleaner}
			\begin{tabular}{ccc|ccc}
				\toprule
				\textbf{TF-IDF} & \textbf{Anti.} & \textbf{Dup.} &  HDFS & BGL & Thunderbird\\
				\midrule
				&  &  & 0.0\% & 0.0\% & 0.0\%\\
				$\checkmark$ &  &  & 0.0\% & 21.65\% & 5.19\%\\
				$\checkmark$ & $\checkmark$ &  & 6.90\%  & 44.33\% & 35.34\%\\
				$\checkmark$ & $\checkmark$ & $\checkmark$ & 48.28\%  & 73.13\% & 69.91\%\\
				\bottomrule
			\end{tabular}
		\end{table}
		\endgroup
		
		To evaluate the effectiveness of each component, we conduct an ablation study for anomaly detection. We assess under various configurations: utilizing TF-IDF alone, combining TF-IDF with the Anti-Events Optimizer (Anti.), and employing both TF-IDF and Anti-Events Optimizer (Anti.) alongside the Duplicative-Events Separator (Dup.). As presented in table~\ref{tab: ablation-logcleaner}, each component plays a role in reducing the number of events, with the Duplicative-Events Separatorr (Dup.) having the most pronounced effect.
		
\begin{figure}[htbp]
			\centering
			\subfigure[BGL]{
				\begin{minipage}{0.46\linewidth}
					\centering
					\includegraphics[width=\textwidth]{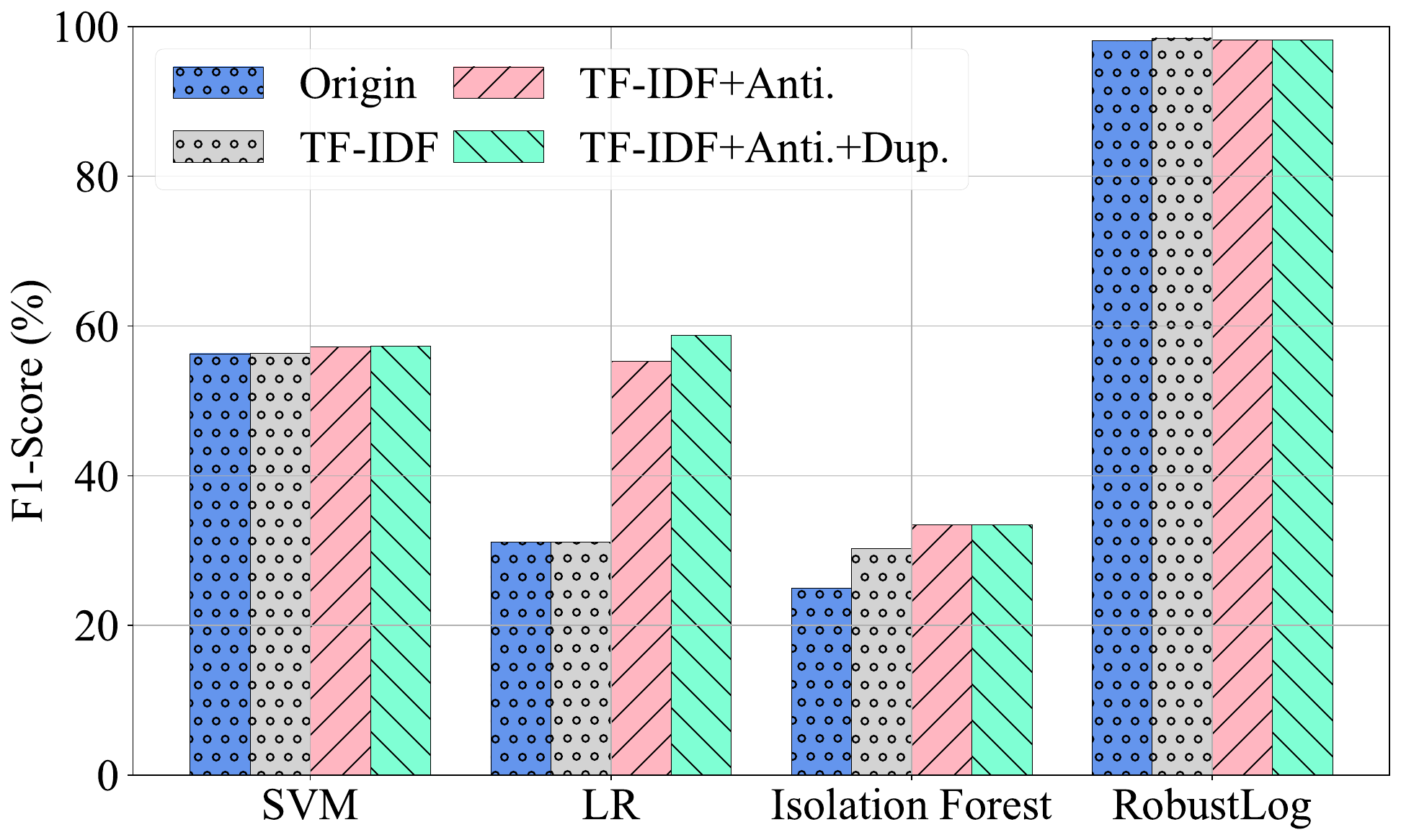}
					\label{fig: ablation-bgl}
				\end{minipage}
			}
			\subfigure[Thunderbird]{
				\begin{minipage}{0.46\linewidth}
					\centering   
					\includegraphics[width=\textwidth]{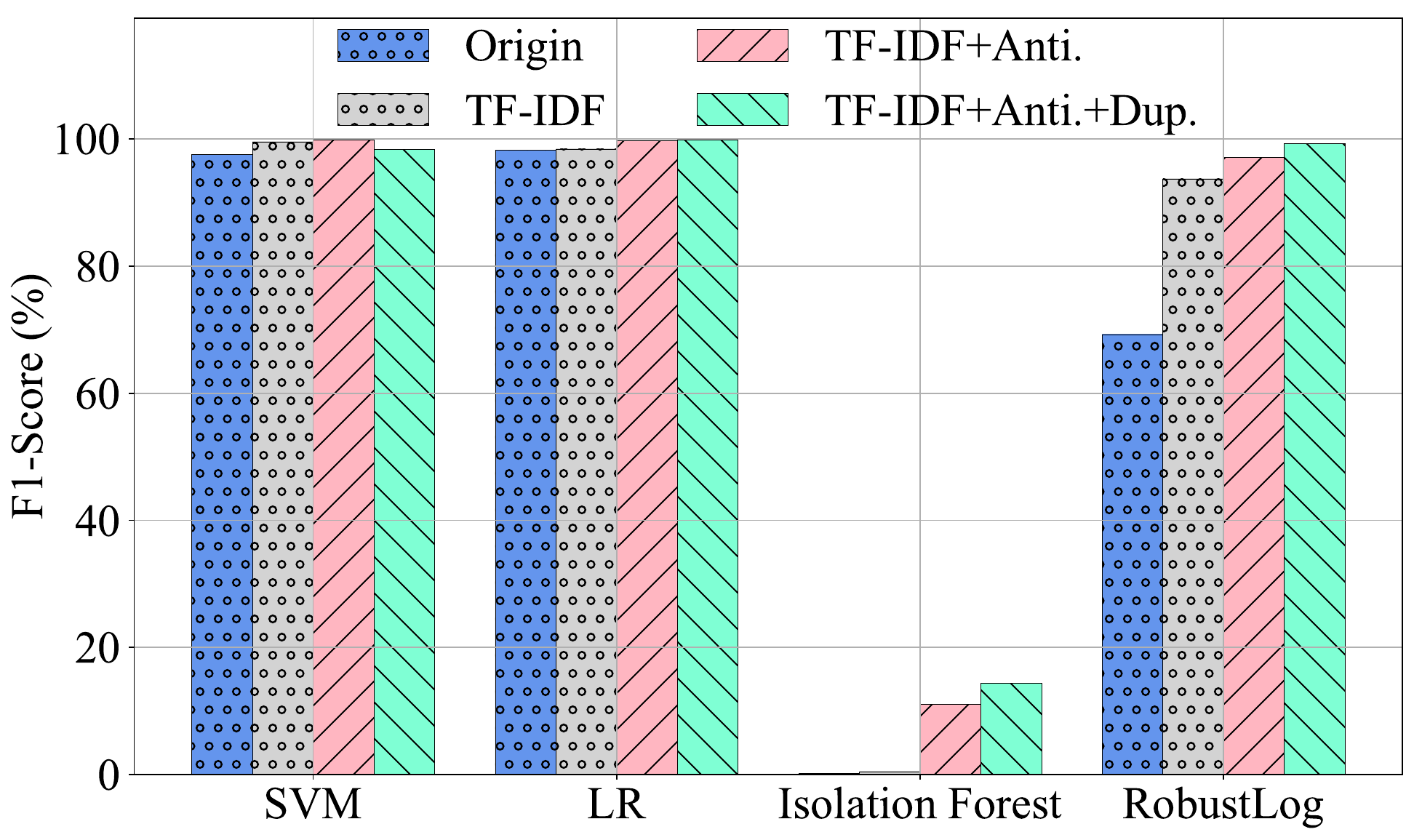}
					\label{fig: ablation-thunderbird}
				\end{minipage}
			}
			\caption{Ablation Experiment of F1-score}
			\label{fig: ablation}
		\end{figure}

		We also validate the performance changes of models on the BGL and Thunderbird datasets. As illustrated in figure~\ref{fig: ablation}, in some instances, the model's performance remains unaffected with the addition of more components. However, for models such as LR and Isolation Forest in the BGL dataset and RobustLog in the Thunderbird dataset, the effectiveness of the models increases with the addition of components. Notably, the Anti-Event Optimizer (Anti.) plays the most significant role in this improvement.
		
		\subsection{Influence of Hyperparameters}
		
		\begin{figure}[htbp]
			\centering
			\subfigure[HDFS]{
				\begin{minipage}{0.46\linewidth}
					\centering
					\includegraphics[width=\textwidth]{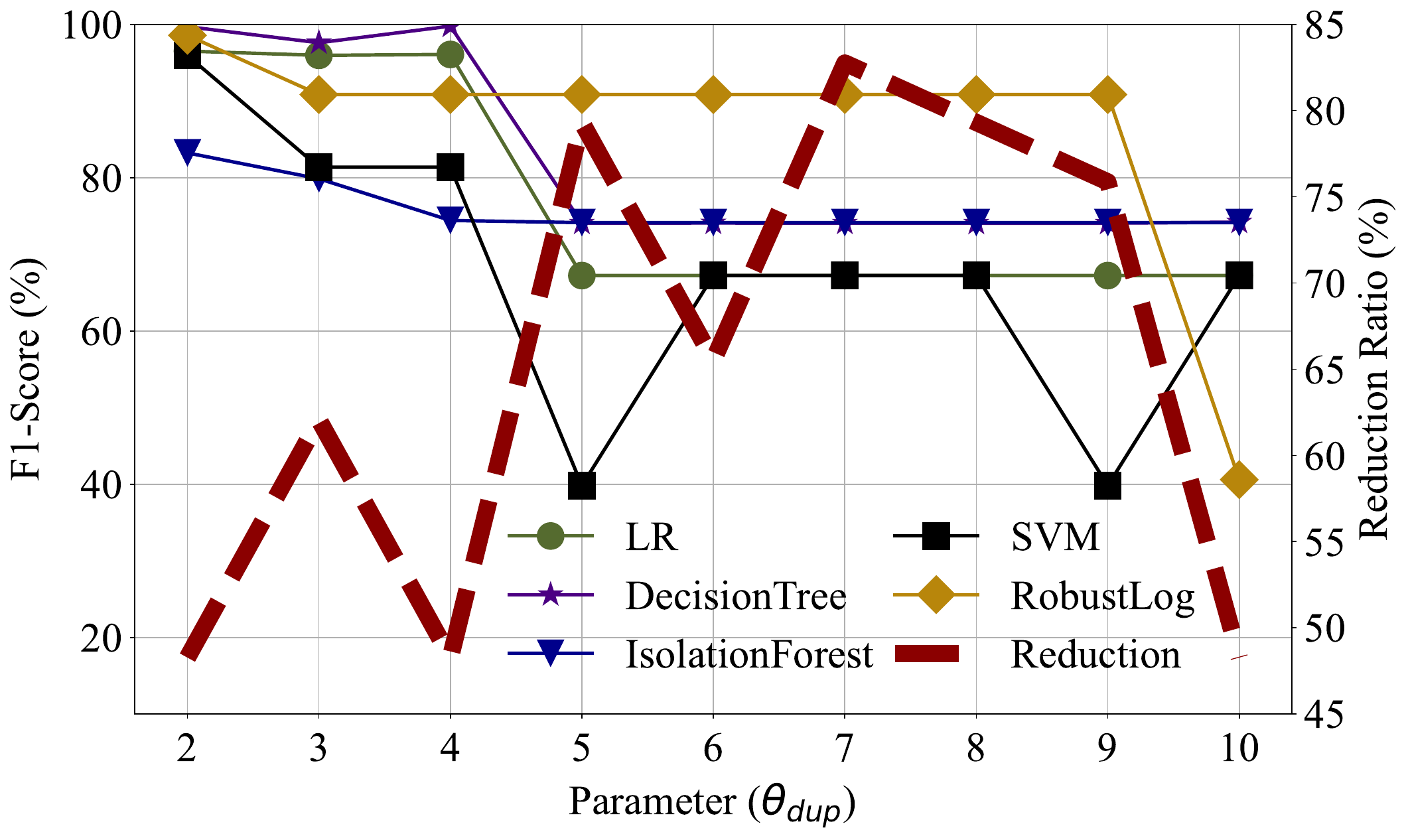}
					\label{fig: min_samples-hdfs}
				\end{minipage}
			}
			\subfigure[BGL]{
				\begin{minipage}{0.46\linewidth}
					\centering   
					\includegraphics[width=\textwidth]{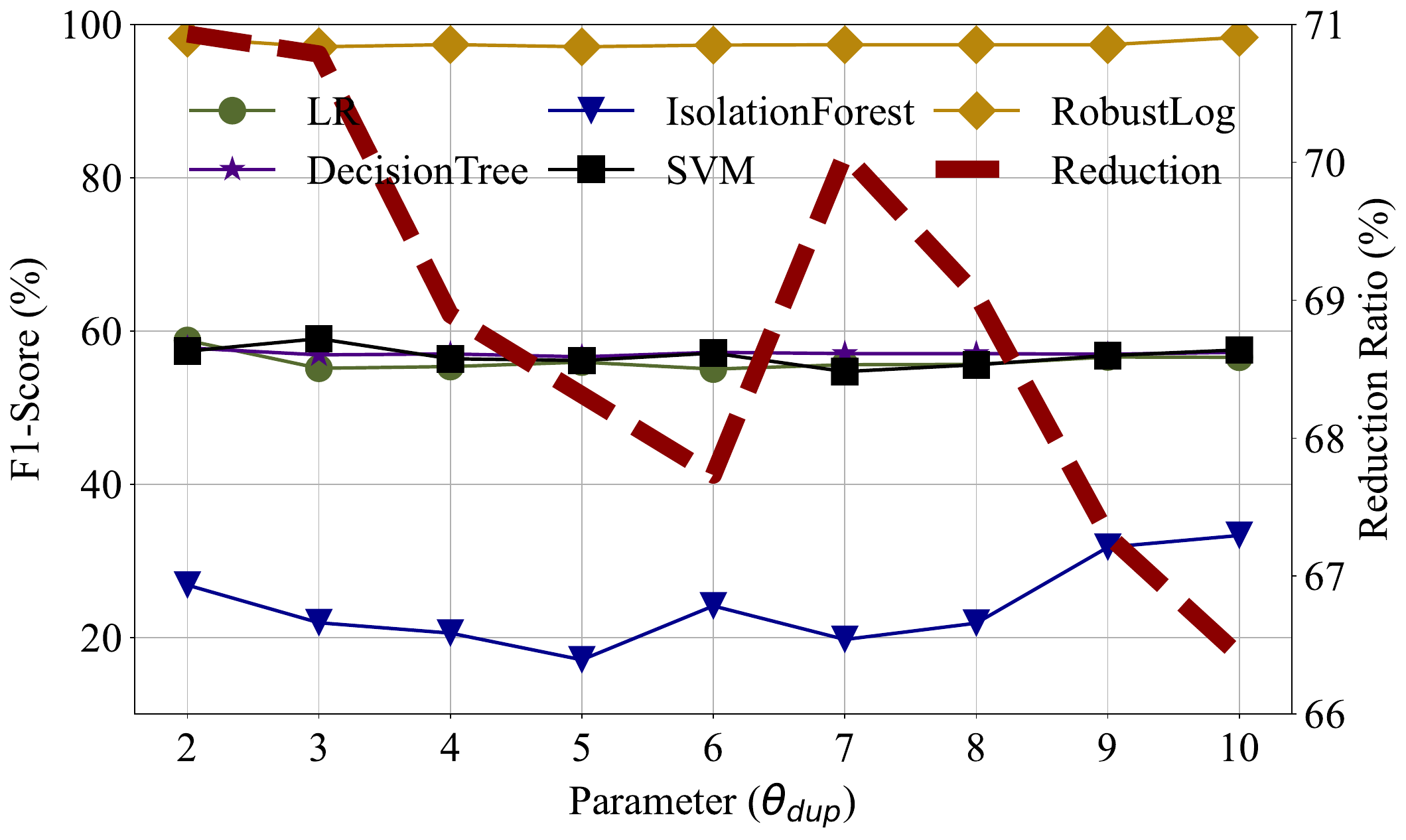}
					\label{fig: min_samples-bgl}
				\end{minipage}
			}
			\caption{Evaluation Results by Varying $\theta_{dep}$}
			\label{fig: min_samples}
		\end{figure}
		
		To verify whether LogCleaner have chosen the optimal hyperparameters, experiments by varying the hyperparameters are carried out. As depicted in figure~\ref{fig: min_samples}, it's evident that while in HDFS, as $\theta_{dep}$ increases, the number of event reductions rises, the model's performance diminishes. Conversely, in BGL, as $\theta_{dep}$ increases, the model's performance remains consistent, but the number of event reductions drastically decreases. Thus, a setting of $\theta_{dep}=2$ is a relatively optimal parameter.
		
		\begin{figure}[htbp]
			\centering
			\subfigure[HDFS]{
				\begin{minipage}{0.46\linewidth}
					\centering
					\includegraphics[width=\textwidth]{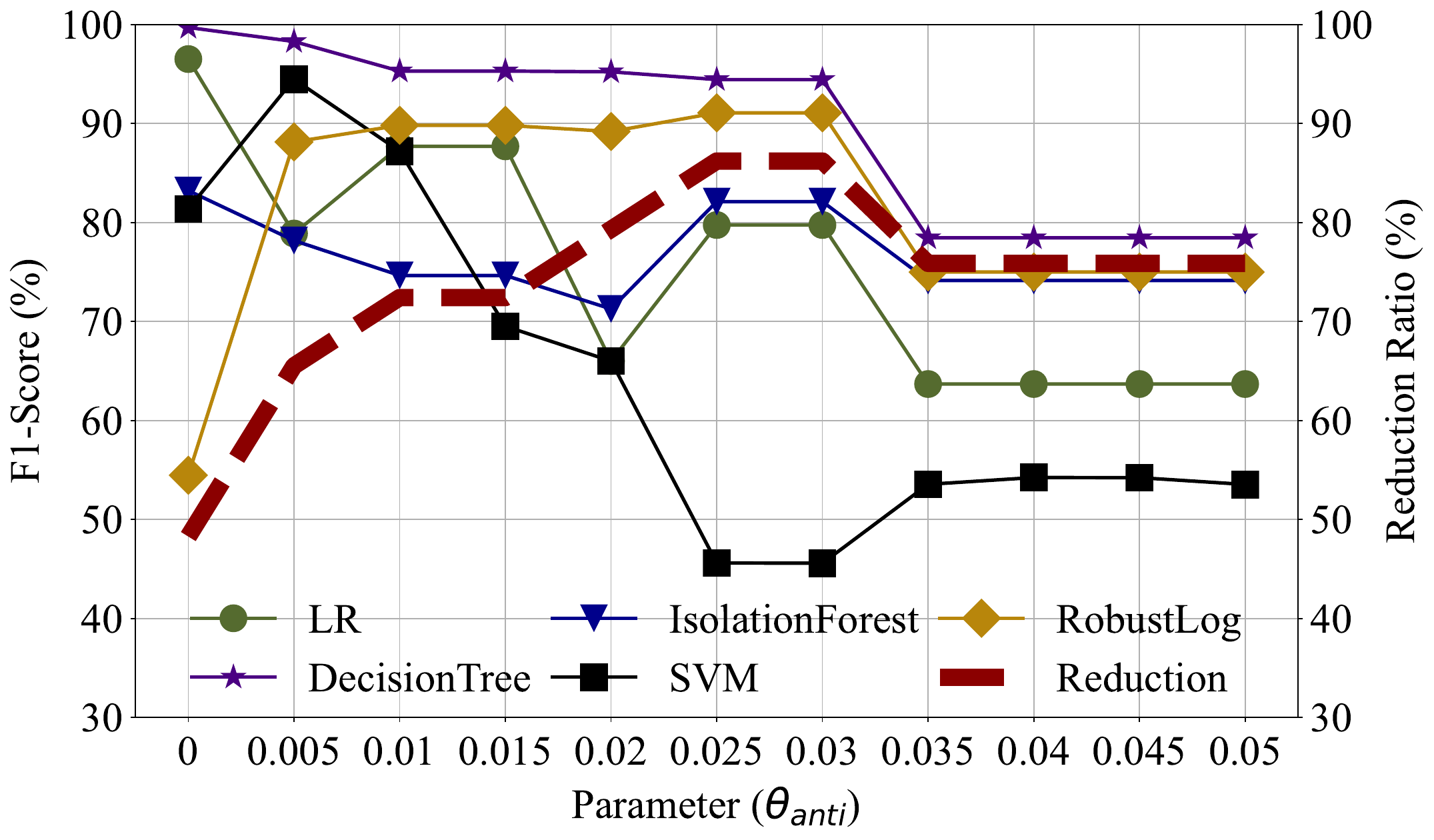}
					\label{fig: mutual_info-hdfs}
				\end{minipage}
			}
			\subfigure[BGL]{
				\begin{minipage}{0.46\linewidth}
					\centering   
					\includegraphics[width=\textwidth]{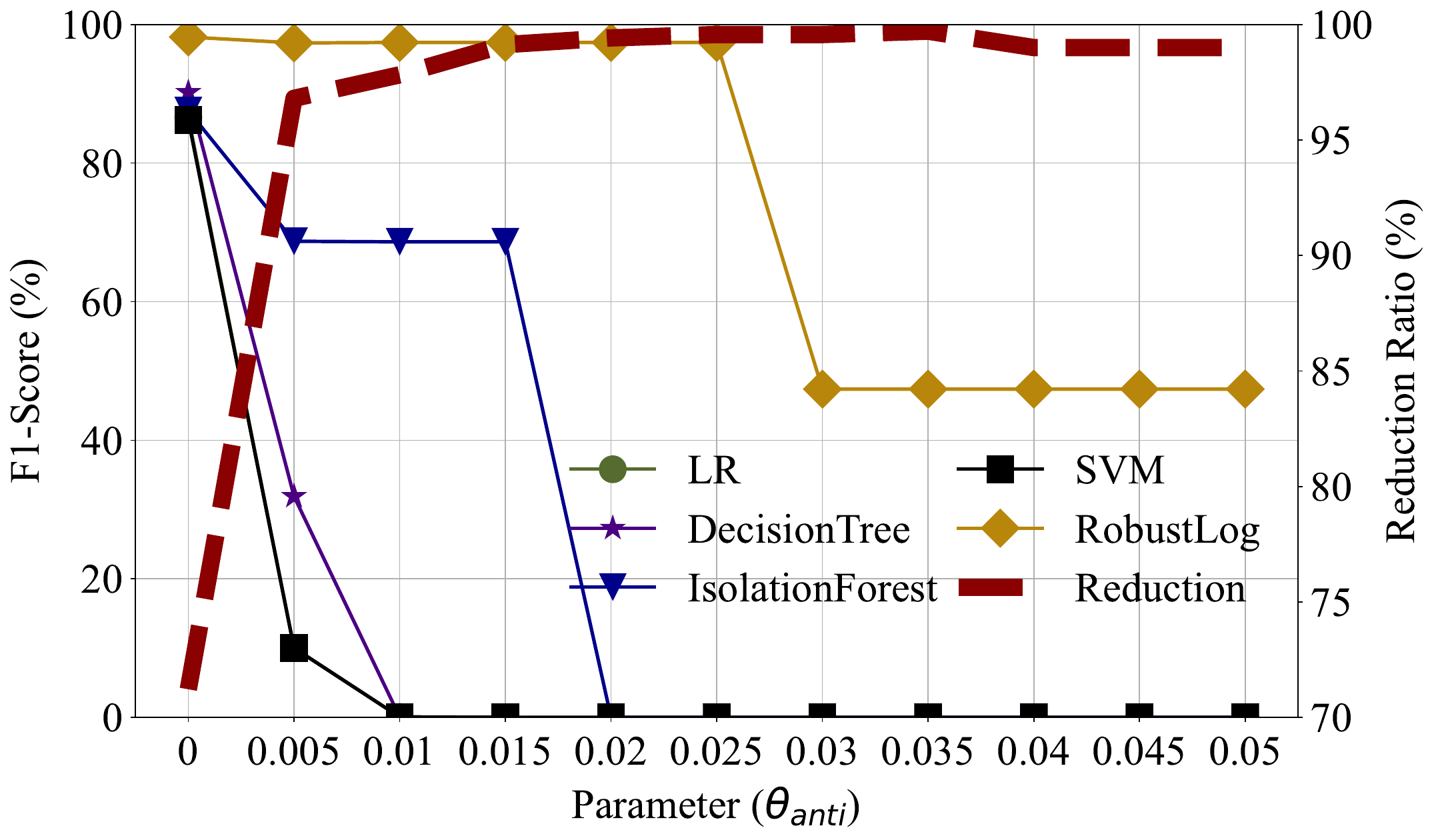}
					\label{fig: mutual_info-bgl}
				\end{minipage}
			}
			\caption{Evaluation Results by Varying $\theta_{anti}$}
			\label{fig: mutual_info}
		\end{figure}
		
		We also conduct experiments related to $\theta_{anti}$. As illustrated in figure~\ref{fig: mutual_info}, for both HDFS and BGL datasets, as $\theta_{anti}$ increases, event reduction grows. However, the performance of various models also deteriorates. Thus, to ensure optimal model effectiveness, it is prudent to set $\theta_{anti}=0$.
		
		\section{Related Work}
		
		\subsection{Log-based Anomaly Detection}
		
		Log analysis for anomaly detection is a well-established research area\cite{du2017deeplog, jia2021logflash, jia2022augmenting, zhang2019robust, meng2019loganomaly, yang2021semi, han2021unsupervised, liu2022uniparser}. These methodologies typically involve extracting templates and key information from logs, followed by constructing models for anomaly detection and classification. There are mainly two types of models in this domain: graph-based and deep-learning models.
		
		Graph-based models leverage log events parsed from log files to create a graph-based representation. They detect conflicts and anomalies by comparing event sequences against this graph. For instance, LogFlash\cite{jia2021logflash} utilizes a real-time streaming process for log transitions, enhancing the speed of anomaly detection. HiLog\cite{jia2022augmenting} performs an empirical study on four anti-patterns that challenge the assumptions underlying anomaly detection models, proposing a human-in-the-loop approach to integrate human expertise into log-based anomaly detection.
		
		Deep-learning models, conversely, use various neural networks to model sequences of log events. LogRobust\cite{zhang2019robust} applies Term Frequency-Inverse Document Frequency (TF-IDF) and word vectorization to convert log events into semantic vectors, thus improving the accuracy of anomaly detection. UniParser\cite{liu2022uniparser} employs a token encoder and a context encoder to learn patterns from log tokens and their adjacent contexts.
		
		\subsection{Log Compression \& Placement} 
		
		Given the substantial volume of logs generated by modern systems, assisting developers in adding appropriate logging statements is a promising research area, as highlighted in prior studies\cite{zhao2017log20, yu2023logreducer}. Log20\cite{zhao2017log20} enhance debugging capabilities by strategically inserting supplementary logging statements into the source code. Concurrently, LogReducer\cite{yu2023logreducer} leverages eBPF to manage logging overhead in performance-critical areas, ensuring that logging remains effective.
		
		However, the process of archiving massive volumes of logs over extended periods can introduce substantial storage overhead. To address this challenge, several studies have focused on log compression techniques to reduce storage requirements. Approaches such as Nanolog\cite{yang2018nanolog} and CLP\cite{rodrigues2021clp} construct dictionaries for fields in logs and replace strings by referencing these dictionaries. Additionally, LogZip\cite{liu2019logzip} and RoughLogs\cite{meinig2019rough} employ sophisticated statistical models to identify and reduce redundancy in logs.
		
		\section{Discussion}
		
		\subsection{Application of LogCleaner}
		
		LogCleaner has been implemented for a subset of users in Apache IoTDB~\cite{zhang2021two, kang2022separation, zhang2024time}, yielding positive feedback. Beyond its application as described in this paper, some users have employed LogCleaner to identify key events and alert developers about unnecessary print statements in the logs that can be removed. Developers can selectively delete these prints to enhance system performance. It has aided developers in discovering that over 50\% of the log print statements in the system are unnecessary. As a result, the performance of Apache IoTDB has improved by approximately 8\%.
		
		\subsection{Threats to Validity}
		
		The major threats to the validity can be identified as following.
		
		\textbf{Limited models.} In the empirical study, we mainly evaluate six representative models that have publicly available source code. In the future, we plan to re-implement more log-based detection models that have not released their source code, based on the descriptions provided in their papers. Subsequently, a larger-scale evaluation will be conduct.
		
		\textbf{Implementation.} We primarily utilize publicly available implementations of the studied models. The implementation of LogCleaner is also based on popular libraries, and three authors have thoroughly reviewed the source code to ensure accuracy and reliability.
		
		\textbf{Limited datasets.} The experiments are conducted on three log datasets. While they are widely used in existing studies on log-based anomaly detection, they may not fully represent all characteristics of log data. In future research, we plan to conduct experiments on additional datasets to cover a broader range of real-world scenarios.
		
		\section{Conclusion and Future Work}
		
		In this paper, we examine event reduction's effect on log-based anomaly detection models. Through empirical study on six models across three datasets, we identify three distinctive log event types that impact model performance differently. Based on these findings, we propose LogCleaner: an efficient methodology for the automatic reduction of log events in the context of anomaly detection. Serving as middleware between software systems and models, LogCleaner continuously updates and filters \textit{anti-events} and \textit{duplicative-events} in the raw generated logs. This approach not only accelerates the model's inference speed but also enhances the effectiveness of model classification.
		
		In future research, we intend to leverage reinforcement learning to enhance the efficacy of log reduction. Furthermore, we also aspire to integrate LLM to pinpoint key events.
		
		\begin{acks}
		This work was supported by the PKU-ZTE Cooperation Research Project.
		\end{acks}
		
		\bibliographystyle{ACM-Reference-Format}
		\balance
		\bibliography{sample-base}
		
	\end{sloppypar}
\end{document}